\documentclass[journal=apchd5,manuscript=article,layout=twocolumn]{achemso}

\usepackage{mathtools}
\usepackage{ulem}
\usepackage{xcolor}

\usepackage{amsmath,amssymb}
\usepackage{soul}

\title{Multi-port programmable silicon photonics using low-loss phase change material Sb$_2$Se$_3$.}

\author{Thomas W. Radford}
\affiliation{School of Physics and Astronomy, University of Southampton, Southampton, SO17 1BJ, United Kingdom}
\email{T.Radford@soton.ac.uk} 

\author{Idris A Ajia}
\affiliation{School of Physics and Astronomy, University of Southampton, Southampton, SO17 1BJ, United Kingdom}

\author{Latif Rozaqi}
\affiliation{School of Physics and Astronomy, University of Southampton, Southampton, SO17 1BJ, United Kingdom}

\author{Priya Deoli}
\affiliation{School of Physics and Astronomy, University of Southampton, Southampton, SO17 1BJ, United Kingdom}

\author{Xingzhao Yan}
\affiliation{Optoelectronics Research Centre, University of Southampton, Southampton, SO17 1BJ, United Kingdom}

\author{Mehdi Banakar}
\affiliation{Optoelectronics Research Centre, University of Southampton, Southampton, SO17 1BJ, United Kingdom}

\author{David J Thomson}
\affiliation{Optoelectronics Research Centre, University of Southampton, Southampton, SO17 1BJ, United Kingdom}

\author{Ioannis Zeimpekis}
\affiliation{School of Electronics and Computer Science, University of Southampton, Southampton, SO17 1BJ, United Kingdom}
\alsoaffiliation{Optoelectronics Research Centre, University of Southampton, Southampton, SO17 1BJ, United Kingdom}

\author{Alberto Politi}
\affiliation{School of Physics and Astronomy, University of Southampton, Southampton, SO17 1BJ, United Kingdom}

\author{Otto L. Muskens}
\affiliation{School of Physics and Astronomy, University of Southampton, Southampton, SO17 1BJ, United Kingdom}
\email{O.muskens@soton.ac.uk}

\begin{document}


\begin{abstract}
Reconfigurable photonic devices are rapidly emerging as a cornerstone of next generation optical technologies, with wide ranging applications in quantum simulation, neuromorphic computing, and large-scale photonic processors. A central challenge in this field is identifying an optimal platform to enable compact, efficient, and scalable reconfigurability. Optical phase-change materials (PCMs) offer a compelling solution by enabling non-volatile, reversible tuning of optical properties, compatible with a wide range of device platforms and current CMOS technologies. In particular, antimony tri-selenide ($\text{Sb}_{2}\text{Se}_{3}$) stands out for its ultra low-loss characteristics at telecommunication wavelengths and its reversible switching. In this work, we present an experimental platform capable of encoding multi-port operations onto the transmission matrix of a compact multimode interferometer architecture on standard 220~nm silicon photonics using \textit{in-silico} designed digital patterns. The multi-port devices are clad with a thin film of $\text{Sb}_{2}\text{Se}_{3}$, which can be optically addressed using direct laser writing to provide local perturbations to the refractive index. A range of multi-port geometries from 2$\times$2 up to 5$\times$5 couplers are demonstrated, achieving simultaneous control of up to 25 matrix elements with programming accuracy of 90\% relative to simulated patterns. Patterned devices remain stable with consistent optical performance across the C-band wavelengths. Our work establishes a pathway towards the development of large scale PCM-based reconfigurable multi-port devices which will allow implementing matrix operations on three orders of magnitude smaller areas than interferometer meshes.
\end{abstract}

\maketitle

The development of the next generation of highly reconfigurable, large-scale photonic architectures offers a critical milestone towards the realization of photonic matrix-vector multiplication accelerators \cite{PerezLopez2025, Zhou2022}, convolutional operators \cite{Meng2023}, optoelectronic neural networks \cite{Xu2024} and microwave processors \cite{PerezLopez2024}. The interface between nanophotonics and machine learning is currently receiving tremendous interest for the promise of achieving new platforms for optical information processing \cite{McMahon2023, Brunner2025, Wu2025, Qian2025}. Integrated photonic devices with addressable transmission matrices represent a foundational element in the advancement of analogue optical computing. Universal matrix decomposition has been well established using meshes of interferometers \cite{reck1994experimental, Shen2017, Perez2020, Bogaerts2020,Chen2023, Bao2023}, yet such architectures typically occupy a large physical footprint even in modern, optimized configurations \cite{Xu2024, Xie2025} which may ultimately require scale up strategies beyond a single die \cite{Seok2019}.

As an alternative to cascaded single-mode devices, the potential of exploiting complex wave interference in inverse designed multimode devices has raised substantial interest \cite{Piggott2015, Shen2015, Hughes2019, MohammadiEstakhri2019, Meng2021, nikkhah2024inverse,Fu2023}. Interference based components such as multimode interferometers (MMIs) may be readily employed for matrix decomposition, leveraging controlled interference to direct optical signals to individual output ports, or for mixing states across several waveguides. These devices may be optimized either during design \cite{nikkhah2024inverse}, creating bespoke geometries for a given application. However, in the absence of reconfigurability, practical deployment remains limited, as most implementations are inherently static. Post-fabrication control of silicon-based MMIs could be achieved through electrical heaters with limited  freedom in designing the transfer function \cite{Rosa2016, Kim2018}, or all-optically using ultrafast induced perturbations requiring high-power laser excitation \cite{bruck2016all}.

Next to electrically controlled active modulation \cite{Chen2023,Shekhar2024,Wu2025}, the development of non-volatile reconfigurable platforms can offer longer-term adaptive functionalities, reduced energy footprint and post-fabrication diversification and trimming in high-volume manufacturing\cite{Bogaerts2020,Shastri2021,ChenACSPhoton2022,Fang2023, Bogaerts2020b}. In particular, low-loss optical phase-change materials (PCMs) offer a promising route to enable light modulation within an ultra-compact footprint \cite{fang2021non, RiosPhotonix2021, PrabhatanIScience2023, Tripathi2023, Huang2025}. Amongst the new generation of low-loss optical PCMs investigated in recent years\cite{Huang2025}, antimony triselenide ($\text{Sb}_{2}\text{Se}_{3}$) has become one of the main materials of choice for the development of reconfigurable silicon photonics owing to its favourable optical properties \cite{delaney2020new, Chen2024}, with integration of these new PCMs into CMOS foundries ongoing \cite{Wei2024,Popescu2025}. Reprogramming of $\text{Sb}_{2}\text{Se}_{3}$ based PCM devices has been demonstrated via electrical \cite{RiosPhotonix2021,Fang2022,Yang2023} or optical actuation \cite{delaney2021nonvolatile,WuSciAdv2024, Wu2025} and endurance of over one million write and reset cycles was shown using carefully selected parameters \cite{lawson2024optical,Alam2024}. 

In this article, we report the active reconfiguration of multimode, multi-port devices and demonstrate the programming of permutation matrices with up to 25 elements. This new capability offers an critical step towards the development of new types of ultra-compact devices for optical information processing. Our work makes use of a standard 220~nm silicon photonics platform combined with 30~nm thin films of $\text{Sb}_{2}\text{Se}_{3}$ to functionalize compact MMI geometries. Using direct laser writing of digital patterns, we demonstrate that combinations of weak scattering perturbations can be employed to constitute a matrix operation within the continuously coupled, multi-port device. Iteratively generated refractive perturbation patterns are used to simultaneously control up to 25 transmission matrix elements of an MMI to achieve matrix operations within a greatly reduced footprint compared to other approaches.

In this work the digital patterns are designed using an iterative optimization algorithm built around a 2.5D variational finite difference time domain (varFDTD) simulation engine. Predicting the pixel pattern required to implement a target transmission matrix is a non-trivial problem without a known analytical solution. More sophisticated alternative inverse design methodologies are available, including topology optimization \cite{bendsoe1989optimal, lu2014topological,Piggott2015,nikkhah2024inverse}, non-topology blackbox optimization including direct binary search and genetic algorithms \cite{Digani2022, Mao2021, Shen2015,Yu2017, Jia2021}, and the application of machine learning and artificial intelligence techniques \cite{radford2025inverse, dinsdale2021deep, Banerji2021}. Data-driven machine-learning based optimization strategies, such as those developed in our earlier studies\cite{radford2025inverse,dinsdale2021deep}, are effective when working with large sets of solutions. For the purpose of this study, only a small number of matrices were evaluated for a diversity of geometries, therefore a direct binary search approach was chosen to predict well-performing solutions. Our optimization process begins with the unperturbed MMI device structure, and a target transmission matrix is defined. The $\text{Sb}_{2}\text{Se}_{3}$ region is initialized in its crystalline state (light pixels in presented figures) and divided into a predefined grid of pixels. A pixel is selected at random, and its state is switched to the amorphous phase (dark pixels in presented figures) The modified device is simulated, and a cost function used to evaluate the differential in transmission between the current transmission matrix and the target. If the new configuration results in improved agreement, the change is retained and another pixel is selected, if the agreement is reduced, the pixel is reverted to its prior state. This process is repeated for N pixels, changing the state between amorphous or crystalline where appropriate in pursuit of a global minimum of the cost function.

Devices were fabricated using the 220~nm UK CORNERSTONE silicon-on-insulator (SOI) platform \cite{Littlejohns2020}. The SOI rib waveguides were etched 120~nm into the top silicon layer of the wafer. A 30~nm thick layer of $\text{Sb}_{2}\text{Se}_{3}$ was sputter coated onto the devices before a lift-off process allowing selective deposition of PCM onto active regions of the devices. Final structures were clad with a 50~nm thin ZnS:SiO$_{2}$ capping layer to prevent oxidation of the PCM and to enhance the switching performance of individual pixels. Prior to $\text{Sb}_{2}\text{Se}_{3}$ deposition, a short argon-ion etch was employed to remove any surface oxide, as well as preserve vertical single-mode operation by embedding the PCM directly onto the waveguide structure. The multimode regions of the 2$\times$2, 3$\times$3 and 4$\times$4 MMI devices under study have lateral dimensions of 6$\times$40~\textmu m$^2$ and feature tapered waveguides at input and outputs which aid to reduce losses during out coupling. The 5$\times$5 MMI is slightly wider to accommodate all the ports, with a dimension of 8$\times$40~\textmu m$^2$. 

Switching of $\text{Sb}_{2}\text{Se}_{3}$ was achieved through direct laser writing using a current-modulated diode laser at 639\,nm wavelength (Vortran Stradus). Focused laser pulses of nanosecond time duration enable high-resolution, spatially selective programming through local amorphization of the PCM with individual pixel size of around 0.7~\textmu m \cite{delaney2021nonvolatile}. Experimentally, this was implemented by focusing the laser using a 50$\times$, 0.42 numerical aperture (NA) long-working distance microscope objective (Mitutoyo Plan APO NIR) positioned above the sample. Reflected light was imaged onto a colour CCD camera allowing top-down imaging of the sample to position the writing spot. An electronic pulse generator (BK precision 4063b ) was used to control the pulse duration and optical power of the laser through a combination of digital and analogue voltage outputs. More information about specific switching parameters can be found in the supporting information. The objective was mounted on a three-axis stage assembly (Newport), which provided accurate positioning of the laser spot across the PCM region to program pixelated patterns of perturbations onto the device area.

\begin{figure*}[h]
    \centering
    \includegraphics[width = 0.9\linewidth]{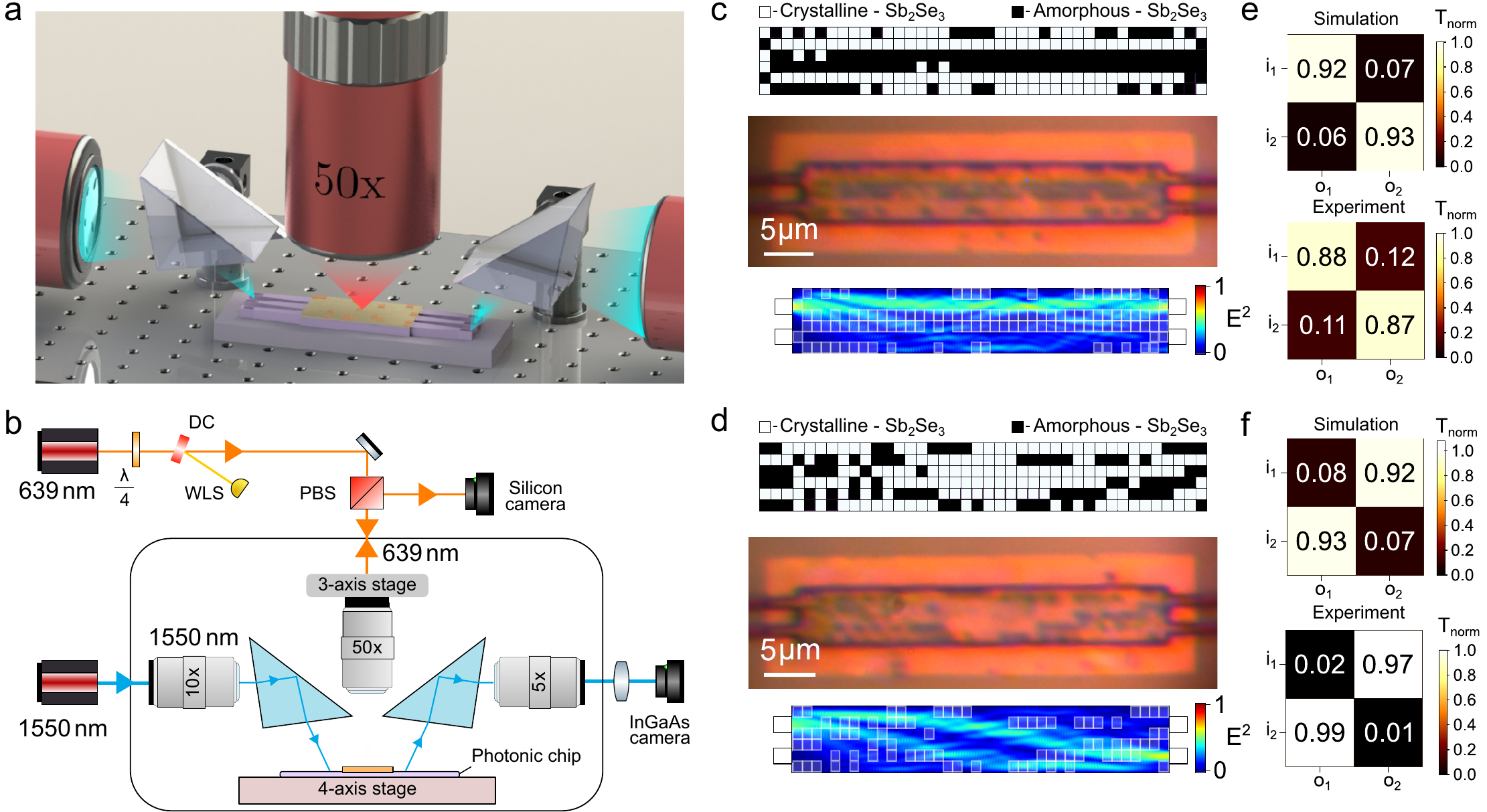}
    \caption{(a) 3D artistic render of the constructed contact free prism coupler used for direct laser writing of PCM films as well as device characterization. (b) Detailed schematic of the experimental setup. (c,d) Experimental results for the bar (c) and cross (d) states of the $2 \times 2$ multi-port switch with designed perturbation maps (top), photograph of MMI after switching (middle), and calculated near-field map ($|\text{E}|^2$) for top input (bottom). (e,f) Simulated (top) and experimental (bottom) transmission matrices for the two states.}
    \label{fig:1}
\end{figure*}

The experiment requires simultaneous access to multiple device ports. However, space for in- and out-coupling using conventional multi-fibre arrays was restricted due to the working distance of the direct-write optical microscope. Therefore, a customized setup was built using a pair of free-space non-contact prisms for coupling in and out of the devices. A 3D render outlining the complete experimental process is shown in Figure\ref{fig:1}(a) with an experimental schematic provided in Figure~\ref{fig:1}(b) and more detail outlined in the supporting information. The two right angle optical prisms were positioned above the chip at a slight tilt to allow coupling at the optimal grating angle of 10$^\circ$. A fibre coupled step-tunable laser source (Keysight N7779C) was focused through the input prism using a 10$\times$ microscope objective and coupled into a single input waveguide of the device. On the output side, multiple grating couplers were imaged simultaneously, collecting the output signal via a separate 5$\times$ objective which projects the image of the output gratings onto an InGaAs camera (Raptor Photonics Owl-320Hs). Areas of interest corresponding to individual grating outputs were selected to quantify the relative intensity distribution across the output ports using post-processing in Python. A more detailed overview of the data collection methodology is given in the Supporting Information.

As a first demonstration of the experimental capability, we investigated the switching of a 2$\times$2 MMI splitter. In the unperturbed state (crystalline background), the device geometry was not optimized for any particular self-imaging condition and showed significant out of plane scattering of input light and insertion loss exceeding 7.5~dB. Programming of the device resulted in the concerted action of the perturbations guiding light towards the targeted output waveguides. The chosen matrices implement an identity transformation (bar state, Figure~\ref{fig:1}c,e) and its inverse (cross state, Figure~\ref{fig:1}d,f). The simulated near field maps for the top input show that the designed pattern efficiently guides light to the corresponding output, similar simulation results are obtained for the other port (Supporting Information). In the simulation, a splitting ratio of around 92\%:8\% was obtained for both states, or a port extinction of 10.7~dB. Due to the spatial separation in optical path for the identity transformation, light may be routed with low crosstalk by selecting a geometry which switches straight paths of PCM to the amorphous state, effectively forming low-index channels that guide light akin to waveguides \cite{shixin2025reconfigurable,Shang24}. 

Both experimental patterned devices exhibit strong agreement with simulated transmission behaviour, with minor deviations likely attributable to differences in perturbation strength between simulation and experiment as well as inaccuracies in the laser writing process and alignment. The cross state outperforms the simulated pattern, achieving 15 dB port extinction. While the identity (bar) and cross configurations of the 2$\times$2 matrix form a relatively simple target, it serves as a clear validation of the platform’s capability. In the following we place greater emphasis on the design and analysis of larger-scale devices, which offer a broader design space of non-trivial matrix operations. 

\begin{figure*}[h!]
    \centering
    \includegraphics[width = 0.8\linewidth]{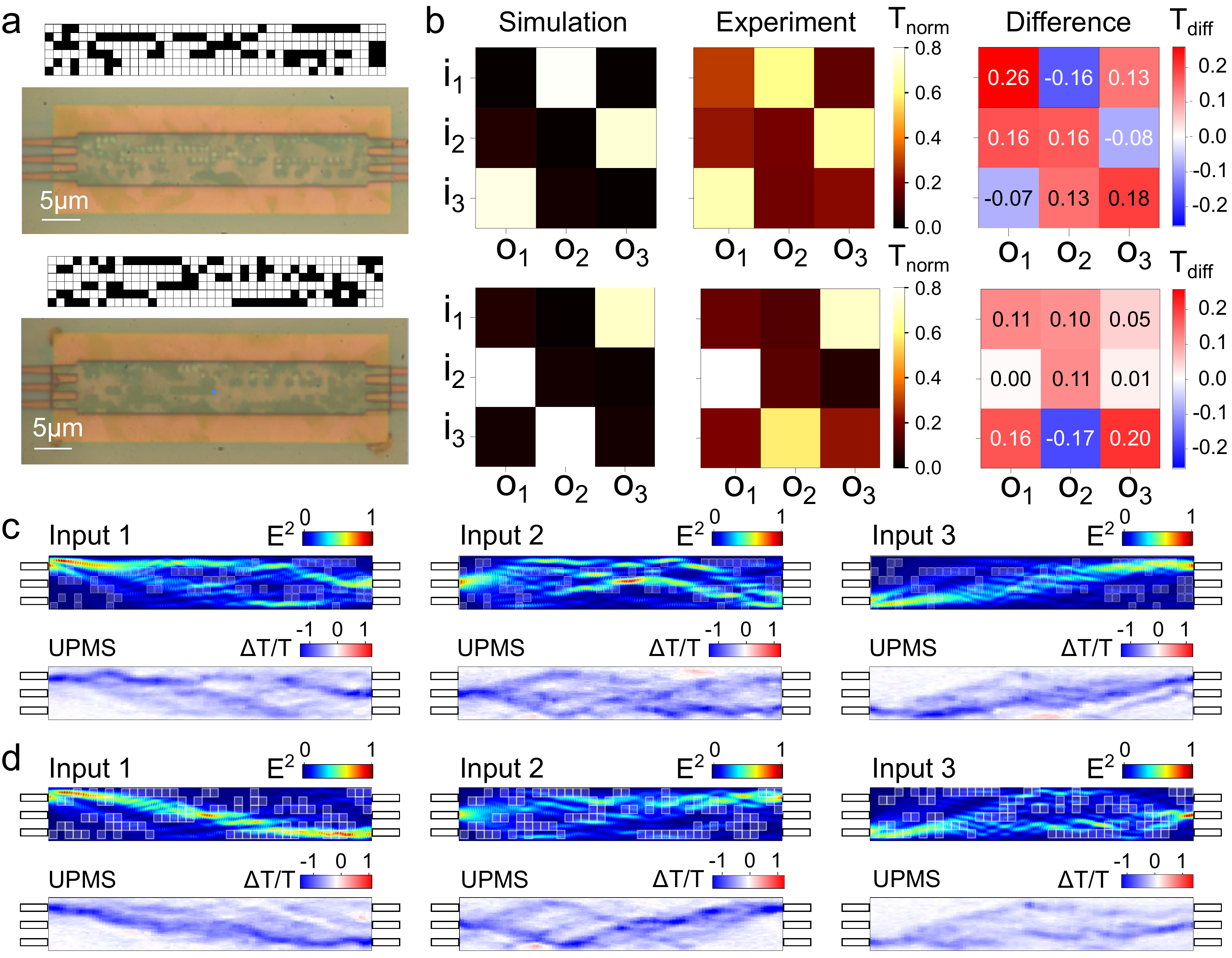}
    \caption{(a) Programmed 3$\times$3 MMI device using patterns designed to implement two of the available permutation matrices. Recorded transmission matrices (b) show good agreement with simulated performance demonstrating average programming errors of $\sim$12\% and an average cosine similarity in excess of 0.9. UPMS measurements (c-d) are used to validate simulation models (top) by probing the electric field distribution within patterned devices.}
    \label{3x3_programmed_mmi_device}
\end{figure*}

Larger multi-port devices offer a broader range of potential applications, such as more powerful analogue simulation of complex systems \cite{nikkhah2024inverse}. However, increasing the number of ports also significantly increases the complexity of the designed pattern required to obtain a target transmission target. Figure~\ref{3x3_programmed_mmi_device} demonstrates the programming of permutation matrices using a 3$\times$3 MMI. The 3$\times$3 coupler has six available permutation matrices, which are orthogonal matrices and correspond to the symmetric group $S_3$. Corresponding designed pixel patterns are shown in Figure~\ref{3x3_programmed_mmi_device}(a) together with photographs of the experimentally written patterns. Due to the increased complexity of matching nine different matrix elements, it becomes more challenging to achieve the desired response with a single pixel pattern. Figure~\ref{3x3_programmed_mmi_device}(b) shows resulting transmission matrices for both the simulation and the experimental realization. The simulations achieve agreement to within 92\% of the ideal permutation matrix, as defined by the cosine similarity of each port, and optimized port transmissions exceeding 80\% (-0.9 dB), matching previous more in-depth modelling efforts for a similar device geometry \cite{dinsdale2021deep}. Experimentally, port transmissions exceeding 75\% are achieved for all ports. Other non-optimized ports show a non-zero background intensity, which is attributed to imperfections in routing the signal to the output through the multiple interference of weakly scattered light paths. The difference matrix for each case is also shown in Figure~\ref{3x3_programmed_mmi_device}(b), which represents the experimental transmission minus the simulated transmission. All experimental matrix elements are normalized to a straight waveguide reference device without PCM and small changes in absolute device efficiency could cause an inaccuracy of the overall transmission as indicated by consistently positive port sums in this work.

To complement our transmission matrix measurements and gain more insight in the flow patterns generated by the perturbed MMIs, we performed Ultrafast Photomodulation Spectroscopy (UPMS) mapping. UPMS is an ultrafast pump-probe techniques developed in our previous work \cite{Vynck2018, bruck2015device} which can be used to map the local intensity distribution in programmed devices \cite{blundell2025ultracompact, dinsdale2021deep}. In short, the devices were transferred to a different fibre-coupled setup and individual port combinations were measured in a pump and probe configuration using an ultrafast pulsed laser at 1560 nm wavelength (Menlo C-Fibre) as the probe and a frequency quadrupled 390~nm UV laser as the pump. The pump was modulated at 5 MHz using an acousto-optic modulator (AA Opto-Electronic) and focused onto the surface of the MMIs by a 50$\times$ objective with an NA of 0.55 (Mitutoyo). A Thorlabs dual-axis galvo scanner with dielectric mirrors was used to scan the pump signal across the active area of the MMI. The change in probe signal due to the perturbation induced by the pump laser was recorded using a lock-in amplifier (Zurich Instruments). Figure~\ref{3x3_programmed_mmi_device}(c) and (d) show maps of the UPMS differential transmission response obtained by summing the individual signals from three output ports. These port-summed maps can be directly compared with the local intensity inside the device \cite{Vynck2018}. Corresponding device simulations are presented in the same figures and provide a good qualitative agreement with the observed flow patterns and routing of the signal toward the different outputs. The UPMS maps also reveal some of the parasitic paths routing light toward undesired ports. Overall the experimental maps confirm the operating principle of the programmable MMIs whereby complex flows of light are being produced by multiple weak scattering events in the digital perturbation pattern.

\begin{figure*}[h!]
    \centering
    \includegraphics[width = 0.99\linewidth]{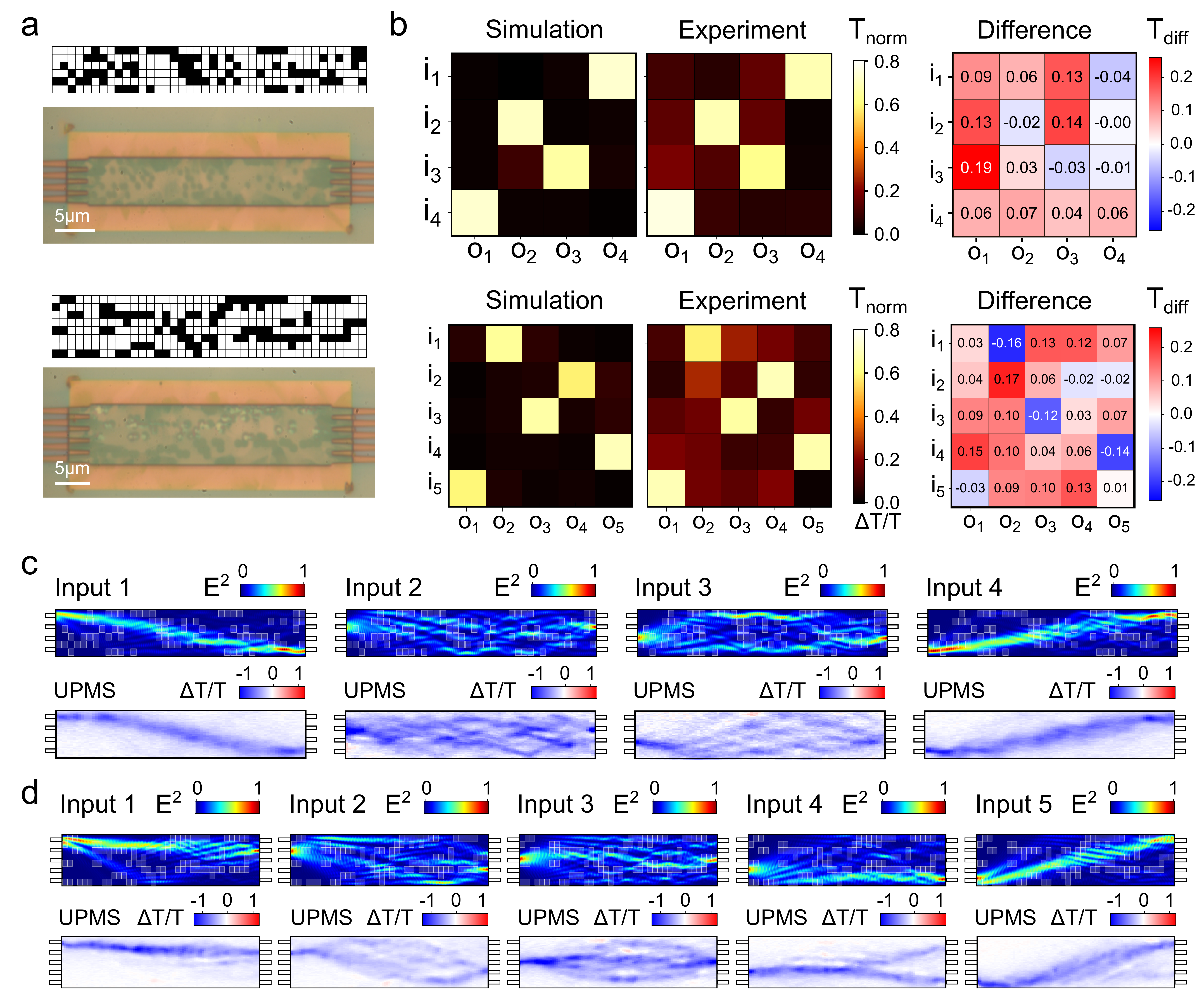}
    \caption{Programming of larger 4$\times$4 and 5$\times$5 MMI devices. Iteratively designed pixel patterns are written into a PCM film on the device, allowing implementation of an orthogonal permutation matrix. The programming error shown is the difference between simulation and recorded performance, indicating slight programming defects that occur during writing. UPMS maps further validate simulation models and indicate the distribution of electric field within the programmed device.}
    \label{4_and_5_port_programmed_mmi}
\end{figure*}

Provided a sufficient ability to achieve suitable design predictions, the technique should be able to program arbitrarily large devices with a comparable degree of accuracy. To test our capability, we attempted programming of 4$\times$4 and 5$\times$5 MMI geometries where we extend the number of degrees of freedom to up to 25 matrix elements, of which in the permutation matrices five elements are designed to non-zero and 20 are designed to be as close to zero as possible. We selected one target design for each device geometry, as presented in  Figure\:\ref{4_and_5_port_programmed_mmi}(a) and (b) representing the available permutation matrices of respectively 24 ($4!$) and 120 ($5!$) for the two geometries. The experimental matrices show overall agreement with the simulation, again achieving average cosine similarities of 0.94, showing the capability of programming larger devices with up to 5$\times$5 ports. Selected output ports show overall high transmission exceeding 75\%, and the background intensity of the other ports is reduced as the residual scattering is distributed over a larger number of ports. Insertion losses of the device (summed over all output ports) compared to a straight waveguide are modest at around 1 dB and do not depend significantly on which input, indicating that all inputs perceive nominally the same losses in the device. UPMS maps confirm the good agreement, where in particular the permutation between top and bottom ports shows accurate paths with little deviation from the design. We observe that the middle ports are more susceptible to deviations, as the interference patterns required are intricate and light get scattered more easily into neighbouring ports.  

\begin{figure*}[h!]
    \centering
    \includegraphics[width = \linewidth]{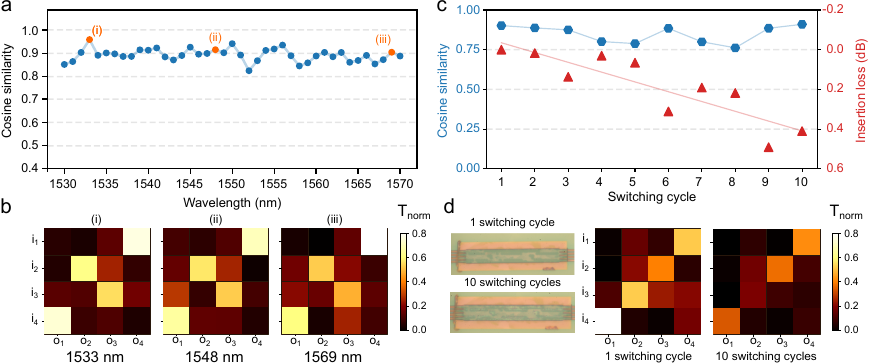}
    \caption{(a) Spectral response across C-band wavelengths of a patterned 4$\times$4 MMI as outlined in Figure\:\ref{4_and_5_port_programmed_mmi}. A selection of matrices are highlighted (b) from which consistent optical performance is recorded. A further pattern to implement in the inverse identity transformation is repeatedly programmed and erased using direct laser writing for both transitions (c-d) outlining the repeated reconfigurability of the presented platform, which exhibits consistent optical performance while imposing small a small insertion loss likely due to damage to the cladding material.}
    \label{Spectra_patterned_device}
\end{figure*}

In applications of reconfigurable integrated components, it is foreseeable that devices may be required to operate across a range of input wavelengths and under varying experimental conditions. In previous studies, patterned devices with weak scattering perturbations have been demonstrated to show broadband characteristics over the telecommunications band \cite{nikkhah2024inverse,Wu2025}. In Figure~\ref{Spectra_patterned_device} we demonstrate the spectral stability of a typical patterned device. Transmission matrices were recorded at wavelengths across the telecommunications C-band range from 1530~nm - 1570~nm. Recorded transmission values were normalized relative to a straight waveguide at the same test wavelength to compensate for the spectral response of the measurement system and grating couplers.

Figure\:\ref{Spectra_patterned_device} plots the cosine similarity between the simulated ''ideal'' device performance at 1550~nm and the recorded transmission matrix at 1~nm increments. For transmission measurements, this similarity metric is bound between 0 and 1 as negative values for transmission are non-physical in our experimental system. From this figure we observe a primarily flat response with average matrix deviations of only 11\%. A randomly chosen selection of example matrices from across this range are shown in Figure\:\ref{Spectra_patterned_device}(b) from which we demonstrate that targeted ports (i1:o4, i2:o3, i3:o3, i4:o1) remain the dominant source of transmission at all studied wavelengths. Figure~S4 confirms this experimental trend using FDTD simulation of the patterned device, both the total and per-port values remain consistent across all test wavelengths demonstrating broadband compatibility of our devices for a range of experimental deployments.

A further important test is to ensure that a single device geometry is able to withstand multiple switching cycles. To this end Figure\:\ref{Spectra_patterned_device}(c-d) present a 4$\times$4 MMI, programmed using a separate pixel pattern designed to implement an inverse identity transformation. The device is optically programmed and subsequently de-programmed using laser pulses of differing power and duration, allowing study of the transmission characteristics throughout this process. The reported cosine similarity is again taken between simulation of the predicted pattern and the recorded values for each switching cycle. General agreement between simulation and experimental results remains good, however, through repeated switching small defects appear to develop in the capping layer of the material which results in increased insertion losses as the experiment is carried out. These defects can be observed in the microscope images shown after thermal annealing of the sample following the 10 programming runs. Scattered light from the surface, indicates non-uniformity in the capping layer which could be caused by the mechanical expansion and stresses in the film \cite{lawson2024optical}. While our work indicate that repeated switching is possible, further work will need to be directed toward optimization of pulse parameters and mechanical stability of the stack.

 The ability to tune device characteristics enables bespoke device programmability, and also provides a promising route for post-fabrication tuning to mitigate design imperfections and increase device throughput. Nanoscale perturbation patterns can be written across a device within 2–3 minutes, a process currently limited primarily by the speed of the implementation using mechanical stages and pulse generation. Significantly faster writing will be achievable using fast laser scanning and optimized electronics and software. 

Computation scalability of the approach towards larger multi-port systems depends on availability of inverse-design algorithms employed and from the accuracy with which patterns can be written into the PCM film. As the device size increases, two compounding optimization challenges emerge: the number of matrix elements requiring optimization grows quadratically with port count, and the combinatorial pixel space expands exponentially with increasing device area, presenting a clear target for future research. In our studies FDTD simulations were separately validated using experimental UPMS mapping, demonstrating good agreement with our simulation models. Other works have demonstrated scaling of computational design up to 10$\times$10 ports \cite{nikkhah2024inverse}. These available studies and our own work jointly provide evidence of the suitability of the multimode platform for a range of complex programming challenges.

Average programming errors (the difference between simulated and recorded performance) across all presented matrices remain around 10\% on average, however, experimental optimization of non-targeted ports is noticeably poorer than in simulation. This suggests that a significant source of experimental error originates from small deviations between the simulated and the programmed device. In particular, variations in spot size, shape, and pixel placement are of key importance, as even minor changes to the multimode regions of MMI geometries can dramatically alter output coupling due to the reliance on precise positioning of self-imaging points. As device dimensions grow, the importance of accurate programming further increases, as small inaccuracies compound across the larger number of programming sites required to tune the output of larger devices. Future work will therefore prioritise improvements in both the resolving power of the microscope objective and the lateral resolution of the programming stage to more faithfully reproduce simulated device geometries.

The programmed devices generally show low insertion losses compared to a straight waveguide reference and offer broadband compatibility across all C-band wavelengths with minimal deviations in the transmission matrix shape from those recorded at 1550~nm, the wavelength for which the patterns were designed. Simulations presented in supporting information show a consistent theoretical transmission of around 70\% with strong optical contrast between targeted and non-targeted ports for the 4$\times$4 MMI under study. This spectral bandwidth enhances the applicability of the platform for systems requiring multi-wavelength operation or flexible source integration, and can likely be improved upon with the use of more advanced inverse design considering also the spectral bandwidth. Considering the typical density of traditional interferometer mesh components of around 50~mm$^{-2}$ and the typical scaling of components with $N(N-1)/2$ where $N$ is the number of input ports \cite{PerezLopez2025}, the our devices can offer an improvement of more than three orders of magnitude from several mm$^2$ down to $3\times10^{-4}$~mm$^2$ for a 5$\times$5 matrix. In conjunction with the aforementioned improvements in spatial selectivity of programming pixels, this platform therefore could enable compact and reconfigurable implementations of arbitrary unitary transformations, paving the way for programmable photonic circuits in optical logic, analogue computing, and large-scale simulation systems. 

$\text{Sb}_{2}\text{Se}_{3}$-silicon integrated devices offer near-universal programmability of transmission matrices using an ultra-low-loss, passive platform. Once patterned, the device operates without the need for active control or regulation, enabling a variety of emerging optical technologies. Reconfigurable devices utilizing $\text{Sb}_{2}\text{Se}_{3}$ are highly scalable, owing to the minimal footprint and resolution of perturbation sites, limited only by the employed fabrication process and experimental optics chosen. PCM integration enables complex matrix decomposition within a micron-scale, continuously coupled architecture, demonstrating a dramatic reduction compared to conventional approaches that rely on large interferometric meshes which and often require active stabilization. While our current work focuses on permutation operators of intensity, future studies should address the potential for programming of complex unitary operators. This will however require considerably more sophisticated methods for phase measurement at the output ports going beyond our current capabilities. 

In conclusion, we have demonstrated the repeated non-volatile programming capabilities of integrated photonic devices clad with thin films of antimony tri-selenide. Using direct laser writing, we were able to experimentally map a range of target matrix operations onto a number of multi-port MMI devices, highlighting both the versatility and scalability of this approach, but also pointing out some limitations in the current implementation related to imperfections in the coupling and leakage to other ports, as well as in the repeatability studies. The patterned devices exhibit increased coupling efficiency with respect to plain MMIs of the same geometry, while maintaining modest insertion losses of, on average, 2.1~dB even for larger 5-port geometries. Resulting devices allow full control over the amplitude of all available input-output combinations, and achieve average programming accuracy of 90.7\% relative to simulated device models, underscoring their suitability for a broad range of experimental systems. The proposed platform holds strong potential for applications in optical logic, photonic processors, and quantum simulation, where compact, reconfigurable, and power-efficient components are essential.

\section{Data availability}
Supporting data used in this work is openly available from the University of Southampton repository at doi.org/10.5258/SOTON/D3764.

\section{Acknowledgements}
This work was supported financially by EPSRC through grants EP/M015130/1 and EP/W024683/1. Silicon photonic waveguides were manufactured through the UK Cornerstone open access Silicon Photonics rapid prototyping foundry through EPSRC grant EP/L021129/1. D.J.T. acknowledges funding from the Royal Society for his University Research Fellowship. 

\newpage

\bibliography{bibliography}

\end{document}